\documentstyle[11pt]{article}
\begin{document}
\hoffset = -1truecm \voffset = -2truecm
\title{\bf (1+1)-dimensional turbulent and chaotic systems reduced from (2+1)-dimensional
Lax integrable dispersive long wave equation}
\author{Xiao-yan Tang$^{1,3}$, Sen-yue Lou$^{2,1,3}$\thanks{Email: sylou@mail.sjtu.edu.cn} and Ying Zhang$^3$\\
\it \footnotesize \it $^1$Physics Department of Shanghai Jiao Tong University, Shanghai 200030, P. R. China\\
\it \footnotesize \it $^2$CCAST (World Laboratory), P.O. Box 8730, Beijing 100080, P. R. China\\
\footnotesize \it $^3$ Abdus Salam International Centre for
Theoretical Physics, Trieste, Italy}
\date{}

\maketitle

\begin{abstract}
After extending the Clarkson-Kruskal's direct similarity reduction
ansatz to a more general form, one may obtain various new types of
reduction equations. Especially, some lower dimensional turbulent
systems or chaotic systems may be obtained from the general type
of similarity reductions of a higher dimensional Lax integrable
model. Especially, the Kuramoto-Sivashinsky equation and an
arbitrary three order quasi-linear equation which includes the
Korteweg de-Vries Burgers equation and the general Lorenz equation
as two special cases are obtained from the reductions of the
(2+1)-dimensional dispersive long wave equation system. Some types
of periodic and chaotic solutions of the (2+1)-dimensional
dispersive long wave equation system are also discussed.
\end{abstract}
\vskip.1in

\section{Introduction}
To reduce a higher dimensional nonlinear physical model to some
lower dimensional ones is one of the most important approaches in
the study of nonlinear science. Usually one use the standard Lie
group approach to reduce a higher dimensional partial differential
equation (PDE) to lower dimensional ones$\cite{Lie}$. Lately, the
so-called nonclassical Lie group analysis is established to find
lower dimensional similarity reductions$\cite{nonclassical}$. To
find some lower dimensional reductions by using the classical and
nonclassical Lie group approaches, one has to use some tedious
algebraic procedures. In the past decade, to avoid the tedious
algebraic calculation in the finding of the similarity reductions,
a simple direct powerful method is developed$\cite{CK,Lou}$. Using
the direct method, various new similarity reductions of many
physical models are found though these reductions can also be
obtained lately from the nonclassical Lie group
approach$\cite{CK1,Lou1,CK2}$. In $\cite{LouTang}$, the direct
method is extended to find some types of conditional similarity
reductions which have not yet been obtained by means of the
present classical Lie group approach and nonclassical Lie group
approach. In this paper, we try to extend the direct method in
another direction to find lower dimensional reductions that may
not be obtained by using the present classical and nonclassical
Lie group approaches.

In the next section, we discuss the general aspect on the direct
reduction method. In section 3, the (2+1)-dimensional dispersive
long wave equation (DLWE) is used as a concrete example to realize
new reduction idea and to find some new lower dimensional
reductions. In section 4, we use some numerical solutions of the
lower dimensional reduction models to discuss some types of exact
solutions of the (2+1)-dimensional DLWE. The last section is a
short summary and discussion.

\section{General reduction ansatz of direct method}
To reduce many types of (n+1)-dimensional nonlinear PDEs,
\begin{eqnarray}
\Delta(x_i,\ u,\ u_{x_i},\ u_{x_ix_j},\ ...,\ i,j=0,1,...,n)\equiv
\Delta[u]=0, \ (x_0\equiv t),
\end{eqnarray}
it is proven that the special ansatz
\begin{eqnarray}
&&u=\alpha(x_0,x_1,...,x_n)+\beta(x_0,x_1,...,x_n)w(\xi_0,\xi_1,\ ...,\ \xi_{n-1}),\\
&&\xi_i=\xi_(x_0,\ x_1,\ ...,\ x_n),\ \nonumber
\end{eqnarray}
is sufficient instead of
\begin{eqnarray}
u=U(x_0,x_1,...,x_n, w(\xi_0,\xi_1,\ ...,\ \xi_{n-1})),
\end{eqnarray}
where $w(\xi_j)\equiv w(\xi_0,\xi_1,\ ...,\ \xi_{n-1})$ satisfies
an $n$-dimensional PDE. In (2) and (3), $u$ and $w$ may be some
multi-component fields. However, to reduce a higher dimensional
PDE to some lower dimensional ones one may use some more general
ansatzs instead of (3). For instance, the ansatz (3) may be
extended as
\begin{eqnarray}
u=U(x_i,\ w_{\xi_j},\ w_{\xi_{j_1}\xi_{j_2}},\ ...,\
w_{\xi_{j_1}\xi_{j_2}...\xi_{j_k}})\equiv U[w].
\end{eqnarray}
In other words, some types of derivatives of the reduction
function may be included in the primary reduction ansatz. However,
to find some concrete results is quite difficult by using the
general ansatz (4). By using the similar procedure of the
simplification from (3) to (2) for many types of significant
mathematical physics models, one may simplify (4) to
\begin{eqnarray}
u=U_0[w]+U_1[w]w_{\xi_{j_1}\xi_{j_2}...\xi_{j_k}},
\end{eqnarray}
where $w_{\xi_{j_1}\xi_{j_2}...\xi_{j_k}}$ is one of the highest
derivatives of $w$ included in (4) while $U_0[w]$ and $U_1[w]$ are
$w_{\xi_{j_1}\xi_{j_2}...\xi_{j_k}}$ independent.

\section{Special new reductions of the (2+1)-dimensional DLWE}

To give out some concrete results from above general reduction
ansatz, we take the (2+1)-dimensional dispersive long wave
equation (2DDLWE)
\begin{eqnarray}
&& u_{yt}+\eta_{xx}+u_xu_y+uu_{xy}=0,\\
&& \eta_t+u_x +\eta u_x+u\eta_x+u_{xxy}=0
\end{eqnarray}
as a simple example. The equation system (6) and (7) is first
obtained by Boiti \em et al. \rm $\cite{Boiti}$ as a compatibility
condition for a `weak' Lax pair. The infinite dimensional
Kac-Moody-Virasoro type symmetry structure of the model is
revealed by Paquin and Winternitz$\cite{PW}$. The more general
$W_\infty$ symmetry is given in$\cite{Winfty}$. It is proven
that$\cite{PP}$ the 2DDLWE system is fails in passing the
Painlev\'e test both at the WTC's (Weiss-Tabor-Carnevale)
$\cite{WTC}$ meaning and at the ARS's (Ablowitz-Ramani-Segur)
meaning$\cite{ARS}$. Using the special ansatz (2), nine types of
two dimensional similarity reductions and thirteen types of ODE
(ordinary differential equation) reductions has been given by one
of the present authors (Lou)$\cite{dlwe}$.

For simplicity further, we taking the reduction ansatz (5) in a
specific form,
\begin{eqnarray}
&& u=F_1(t,y,w)w_x+F_0(w,w_x,w_{xx})+F_2(w,w_x)w_{xxx},\\
&& v\equiv \eta+1 = u_y= F_{1y}(t,y,w)w_x,\qquad w\equiv w(x,t).
\end{eqnarray}
The reason why we take the ansatz (8) is that we try to find the
reduction equations have the following three order autonomous PDE
form
\begin{eqnarray}
w_t=\alpha w_{xxx}+F_3(w,\ w_x,\ w_{xx})
\end{eqnarray}
for some possible functions $F_3$. The ansatz (9) degenerates two
equations (6) and (7) to a same one.

Substituting (8)-(10) into (6) and/or (7) yields
\begin{eqnarray}
F_{1y}(t,y,w)(F_2(w,w_x)w_x+\alpha)w_{xxxx}+f(t,y,w,w_x,w_{xx},w_{xxx})=0,
\end{eqnarray}
where $f(t,y,w,w_x,w_{xx},w_{xxx})\equiv f $ is a complicated
expression of the indicated variables. Because of $f$ is
$w_{xxxx}$ independent, (11) is valid only for
\begin{eqnarray}
F_2(w,\ w_x)=-\alpha w_x^{-1}.
\end{eqnarray}
Substituting (12) into (11), we have
\begin{eqnarray}
F_{1y}(t,y,w)(F_{3w_{xx}}(w,w_x,w_{xx})+1+w_xF_{0w_{xx}}(w,w_x,w_{xx})
)w_{xxx}+f_1(t,y,w,w_x,w_{xx})=0,
\end{eqnarray}
where $f_1(t,y,w,w_x,w_{xx})\equiv f_1 $ is independent of
$w_{xxx}$. From eq. (13) we immediately have
\begin{eqnarray}
F_3(w,\ w_x,\ w_{xx})=- w_{xx}-w_xF_{0}(w,\ w_x,\ w_{xx})+F_3(w,\
w_x).
\end{eqnarray}
By using Eq. (14), (13) is simplified further to
\begin{eqnarray}
&&(2w_xF_{1y}(t,y,w)F_1(t,y,w)+F_{1y}(t,y,p)F_{3w_x}(w,w_x)+2w_xF_{1yw}(t,y,p))w_{xx}\nonumber\\
&& \qquad +f_2(t,y,w,w_x)=0,
\end{eqnarray}
where $f_2(t,y,w,w_x)\equiv f_2 $ is $w_{xx}$ independent.
Integrating (15) once with respect to $y$, we have
\begin{eqnarray}
&&\left(w_xF_1(t,y,w)^2+F_1(t,y,w)F_{3w_x}(w,w_x)+2w_xF_{1w}(t,y,w)+f_3(t,w,w_x)\right) w_{xx}\nonumber\\
&& \qquad +f_2(t,y,w,w_x)=0,
\end{eqnarray}
with $f_3(t,w,w_x)$ being an integrating function. Because of the
$w_x$ independence of $F_1(t,y,w)$, by vanishing the first term of
(16), we get
\begin{eqnarray}
F_3(w,w_x)=F_{32}(w)w_x^2+F_{30}(w),\qquad
f_4(t,w,w_x)=w_xF_4(t,w),
\end{eqnarray}
and
\begin{eqnarray}
2F_{1w}(t,y,w)+F_1(t,y,w)^2+2F_{32}(w)F_1(t,y,w)+F_4(t,w)=0.
\end{eqnarray}
Because of (17) and (18), Eq. (16) is simplified finally to
\begin{eqnarray}
F_{1t}(t,y,w)+F_{1}(t,y,w)F_{30w}(w)-\frac12F_{30}(w)(F_1(t,y,w)+2F_{32}(w))F_1(t,y,w)+F_5(t,w)=0.
\end{eqnarray}
The compatibility condition of (18) and (19) requires that
\begin{eqnarray}
F_{5}(t,w)=-F_{30ww}(w)+F_{32}(w)F_{30w}(w)-\frac12F_4(t,w)F_{30}(w)+F_{32w}(w)F_{30}(w)
\end{eqnarray}
and
\begin{eqnarray}
&&F_4(t,w)F_{30w}(w)-F_{32}(w)F_{30}(w)F_{32w}(w)-2F_{30w}(w)F_{32w}(w)+\frac12F_{30}(w)F_{4w}(t,w)\nonumber\\
&&-F_{32ww}(w)F_{30}(w)
-F_{32}(w)^2F_{30w}(w)+F_{30www}(w)+\frac12F_{4t}(t,w)=0.
\end{eqnarray}
Now the final results show us that the 2DDLWE (6) and (7)
possesses the following reduction
\begin{eqnarray}
w_t=\alpha w_{xxx}-w_{xx}-w_xF_{0}(w,\ w_x,\
w_{xx})+F_{32}(w)w_x^2+F_{30}(w)
\end{eqnarray}
with arbitrary functions $F_0(w,\ w_x,\ w_{xx}),\ F_{32}(w)$ and $
F_{30}(w)$ and
\begin{eqnarray}
&&u=-\alpha w_x^{-1}w_{xxx}+F_0(w,\ w_x,\ w_{xx})+ F_1(t,y,w)w_x,\\
&&v=\eta+1=F_{1y}(t,y,w)w_x,
\end{eqnarray}
where $F_1(t,y,w)$ is determined by two compatible Riccati
equations (18) and (19) while $F_4(t,\ w)$ and $F_5(t,\ w)$ are
determined by (20) and (21). The simplest solution of (18)--(21)
reads
\begin{eqnarray}
&&F_4(t,w)=F_{32}(w)=F_5(t,w)+2A_2=0,\\
&&F_{30}(w)=A_2w^2+A_1w+A_0,\\
&&F_{1}(t,y,w)=\frac2{w+q(y,t)},\\
&&q_t(y,t)=A_1q(y,t)-A_0-A_2q(y,t)^2,
\end{eqnarray}
where $A_0,\ A_1,\ A_2$ are arbitrary constants.

From the reduction equation (24), we can see that though the
original 2DDLWE system is Lax integrable, possesses infinitely
many symmetries and abundant multi-soliton structures, there still
exist various nonintegrable lower dimensional reductions because
of the entrance of three arbitrary functions $F_0(w,\ w_x,\
w_{xx}),\ F_{32}(w)$ and $F_{30}$. For instance, if we select
$F_0(w,\ w_x,\ w_{xx}),\ F_{32}(w)$ and $F_{30}(w)$ simply as
\begin{eqnarray}
-w_xF_{0}(w,\ w_x,\ w_{xx})+F_{32}(w)w_x^2+F_{30}(w)= ww_x,
\end{eqnarray}
then (22) becomes the well known KdV-Burgers equation
\begin{eqnarray}
w_t=\alpha w_{xxx}- w_{xx}+ww_x
\end{eqnarray}
which is one of the possible candidate to describe the turbulence
phenomena in fluid physics and plasma physics$\cite{turbulence,
KdVB}$. If the functions $F_0(w,\ w_x,\ w_{xx}),\ F_{32}(w)$ and
$F_{30}$ are fixed to satisfy
\begin{eqnarray}
&&-w_xF_{0w_{xx}}(w,\ w_x,\ w_{xx})+F_{32}(w)w_x^2+F_{30}(w)\nonumber\\
&&\qquad =\frac1w[w_{xx}w_x+(c+1)w_x^2]
-w^2w_x-(b+c)w_{xx}-wc(b-ba+w^2)
\end{eqnarray}
with $\alpha=-1$ and $a,\ b,\ c$ are arbitrary constants, then
(24) becomes a (1+1)-dimensional extension
\begin{eqnarray}
w_t=- w_{xxx}-(b+c+1)w_{xx}+\frac1w[w_{xx}w_x+(c+1)w_x^2]
-w^2w_x-wc(b-ba+w^2)
\end{eqnarray}
of the famous chaotic system, the Lorenz system$\cite{Lorenz}$
\begin{eqnarray}
w_{s}=-c(w-g),\ g_{s}=(a-h)w-g,\ h_{s}=wg-bh.
\end{eqnarray}
Actually, the travelling wave reduction of (32),
$w=w(x+b(c+1)t)\equiv w(s)$, is totally equivalent to the Lorenz
system (33).

In principle, any order of derivatives of $w$ may be included in
the ansatz (5). And some types of more complicated reduction
equations can be obtained. For instance, if we insert a fourth
order derivative $w_{xxxx}$ term into the reduction ansatz, we may
obtain many fourth order (1+1)-dimensional PDE reductions. Here we
list only a special example for the reduction equation has a
famous Kuramoto-Sivashinsky (KS) equation form$\cite{KS}$
\begin{eqnarray}
&&u=\pm \frac{2w_x(a_1+a_3Q)}{a_0+a_1w+a_2Q+a_3wQ}-\frac{(\alpha_3\mp1)w_{xx}}{w_x}+\alpha_1w\nonumber\\
&&\qquad +\frac{1}{w_x}(\alpha_5w_{xxxx}+(a_1c_2+a_3c_0)w^2
 +(c_2a_0+c_1a_1+a_2c_0+\alpha_2)w+c_1a_0),\\
&&\eta=\frac{Q_yp_x(a_3a_0-a_2a_1)}{a_0+a_1p+a_2Q+a_3pQ)^2},\\
&&Q_t=a_0c_0+(c_1a_1+a_2c_0-c_2a_0)Q+(c_1a_3-c_2a_2)Q^2,\\
&&w_t+\alpha_1ww_x+\alpha_2w+\alpha_3w_{xx}+\alpha_4w_{xxxx}=0,
\end{eqnarray}
where $a_i,\ \alpha_i, \ i=0,1,...,5$ are arbitrary constants and
$c_0,\ c_1$ and $c_3$ are arbitrary functions of $t$. Various
interesting properties of the chaotic KS equation (37) have been
studied by many authors, say, $\cite{KS}$ and the references
therein.

In the reduction results (22) and (37), the independent variables
are simply taken as $x$ and $t$. Actually, extending these
independent variables to some more general forms is possible
because the model possesses infinitely many symmetries with some
arbitrary functions$\cite{PW, Winfty}$. For instance, using the
finite transformations given in $\cite{PW}$, all the independent
variables of the systems (22) and (37) are changed to some general
forms naturally.

\section{Special solutions}

Now an interesting question is which kinds of exact solutions can
be obtained from our new reduction equations? In this section, we
write and plot down some interesting exact solutions.

\subsection{Multi-dromion solutions}

If we take $F_0(w,\ w_x,\ w_{xx})$ as
\begin{eqnarray}
F_0(w,\ w_x,\
w_{xx})=6w-w_x^{-1}(w_{xx}-w_x^2F_{32}(w)-F_{30}(w)),
\end{eqnarray}
then we know that the $w$ equation (22) is just the well known KdV
equation
\begin{eqnarray}
w_t=\alpha w_{xxx}-6ww_x.
\end{eqnarray}
Then we can use the N soliton solutions of the (1+1)-dimensional
KdV equation to construct the multi-dromion solutions by taking
\begin{eqnarray}
q(y,t)=a_0+\sum_{n=1}^Na_n\tanh (l_ny-y_n)
\end{eqnarray}
with $a_i,\ i=0,1,...,N$ being arbitrary constants and
$A_0=A_1=A_2=0$. Fig. 1 is a plot of the four dromion solution
with $w$ being two soliton solution of the KdV equation
\begin{eqnarray}
&&w=-2\alpha (\ln \phi)_{xx} ,\\
&&\phi=1+\exp(k_1x+\alpha k_1^3t+x_1)+\exp(k_2x+\alpha
k_2^3t+x_2)\nonumber\\
&&\qquad +\frac{(k_1-k_2)^2}{(k_1+k_2)^2}\exp((k_1+k_2)x+\alpha(
k_1^3+k_2^3)t+x_1+x_2)
\end{eqnarray}
 and
\begin{eqnarray}
q=4+\tanh(l_1y-y_1)+2\tanh(l_2y-y_2)
\end{eqnarray}
while the other constants are fixed as
\begin{eqnarray}
k_1=1,\ k_2=1.1, \ l_1=l_2=1,\ \alpha=-1,\ x_1=-3,\ x_2=3,\
y_1=-3,\ y_2=3.
\end{eqnarray}
The figure (1) and all other figures of this paper are plotted at
time $t=0$.
\input epsf
\begin{figure}
\epsfxsize=7cm \epsfysize=5cm \epsfbox{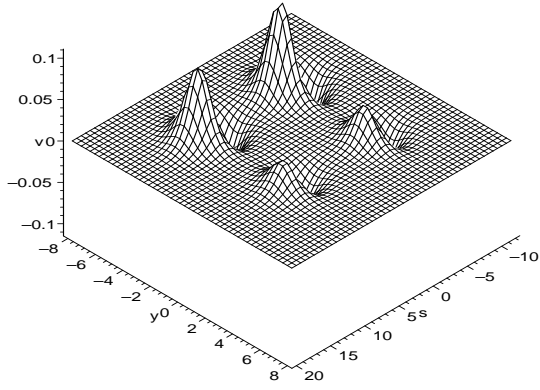}
\caption{\footnotesize Four dromion solution of the 2DDLWE for the
field $v=\eta+1$ given by (24) with (27) and (41)--(44).}
\end{figure}

\subsection{Periodic and chaotic line soliton solutions}

If $q$ is still given by (43) while $w$ is given by (33), then we
may obtain some kinds of periodic or chaotic line soliton
solutions. Because no one has given ever out any exact explicit
solutions of (33), we can only use the numerical solutions of the
generalized Lorenz system to construct exact solutions of the
2DDLWE. For several types of parameter ranges, the solutions of
the Lorenz system are periodic while for other types of parameter
ranges, the solutions of the Lorenz system are chaotic ones. Fig.
2 is a plot of the periodic two line soliton solution of the
2DDLWE with the parameters of the Lorenz system (33) is fixed as
\begin{eqnarray}
a=350 ,\ b=\frac83,\ c=10
\end{eqnarray}
and
\begin{eqnarray}
q=200+\tanh y
\end{eqnarray}
\input epsf
\begin{figure}
\epsfxsize=7cm \epsfysize=5cm \epsfbox{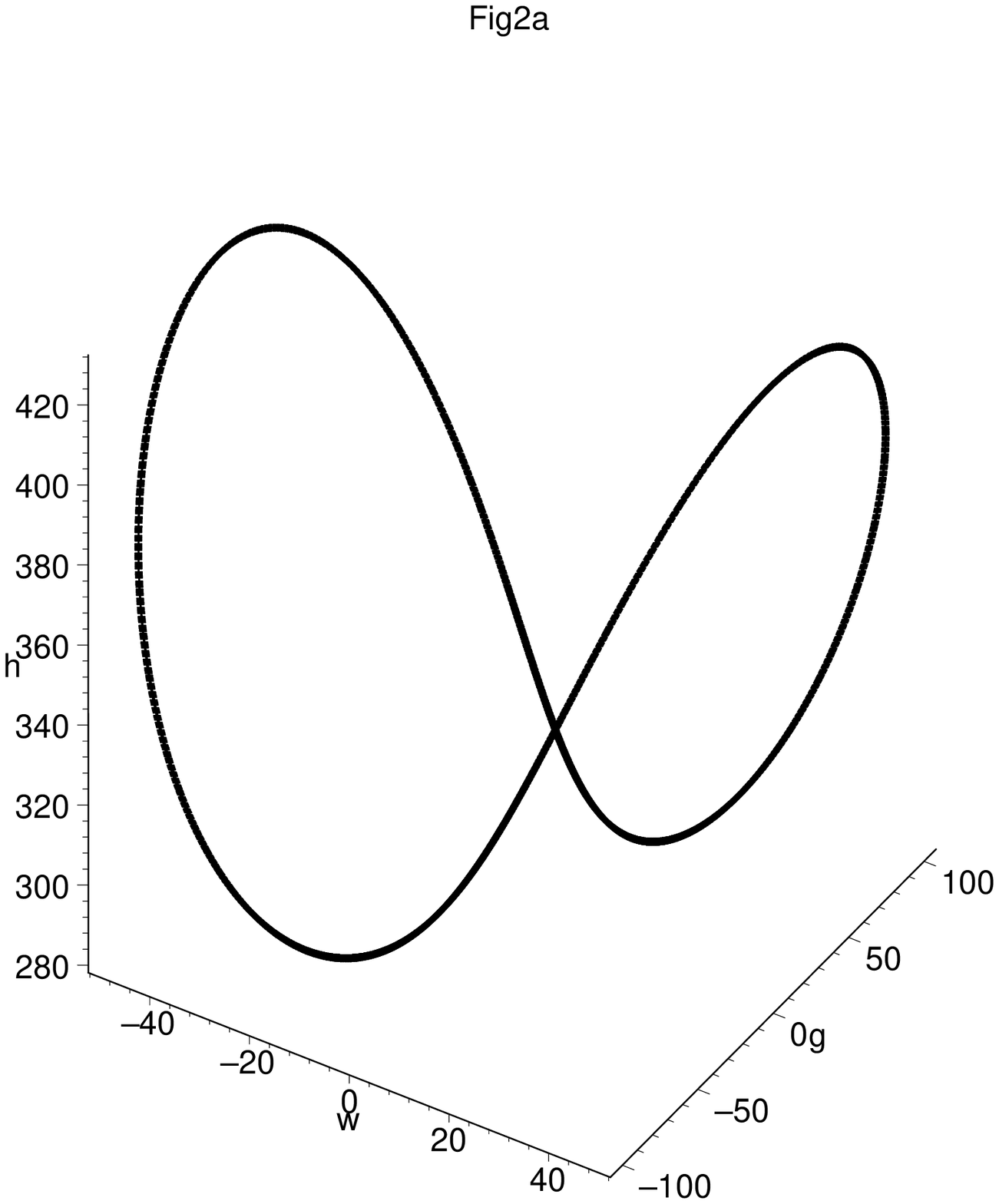}
\epsfxsize=7cm \epsfysize=5cm \epsfbox{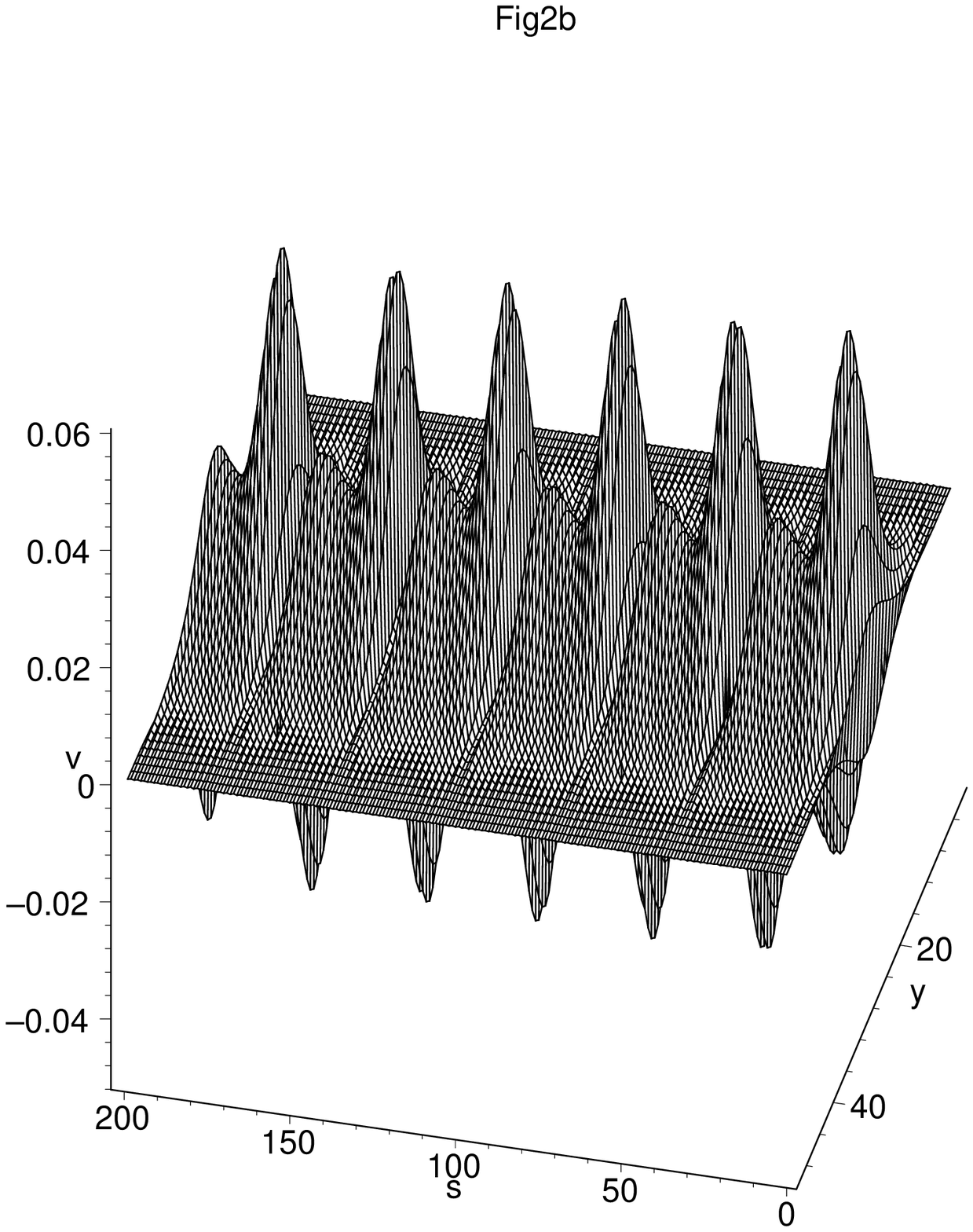}
\caption{\footnotesize (a) Periodic two solution of the Lorenz
system. (b) The related periodic two line soliton solution
 of the 2DDLWE for the
field $v$.}
\end{figure}
From Fig. 2b we can see that the line solution is localized in $y$
direction and periodic in $s(=x+b(c+1)t)$ direction when the
parameters are selected appropriately as (45).

Fig. 3 plots the chaotic line soliton solution of the 2DDLWE with
the parameters of the Lorenz system (33) given by
\begin{eqnarray}
a=60 ,\ b=\frac83,\ c=10
\end{eqnarray}
while the $q$ function still given by (46).
\input epsf
\begin{figure}
\epsfxsize=7cm \epsfysize=5cm \epsfbox{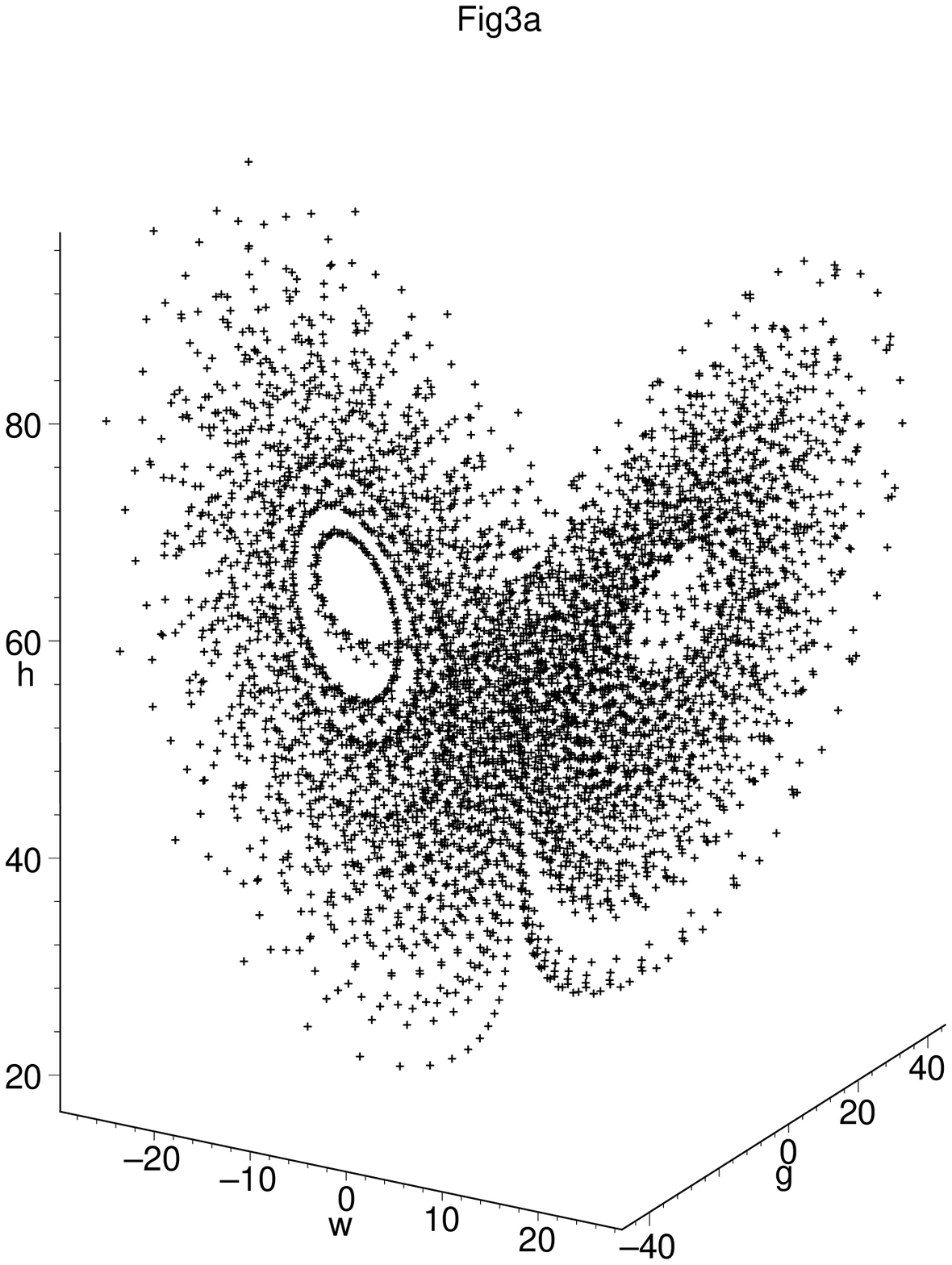}
\epsfxsize=7cm \epsfysize=5cm \epsfbox{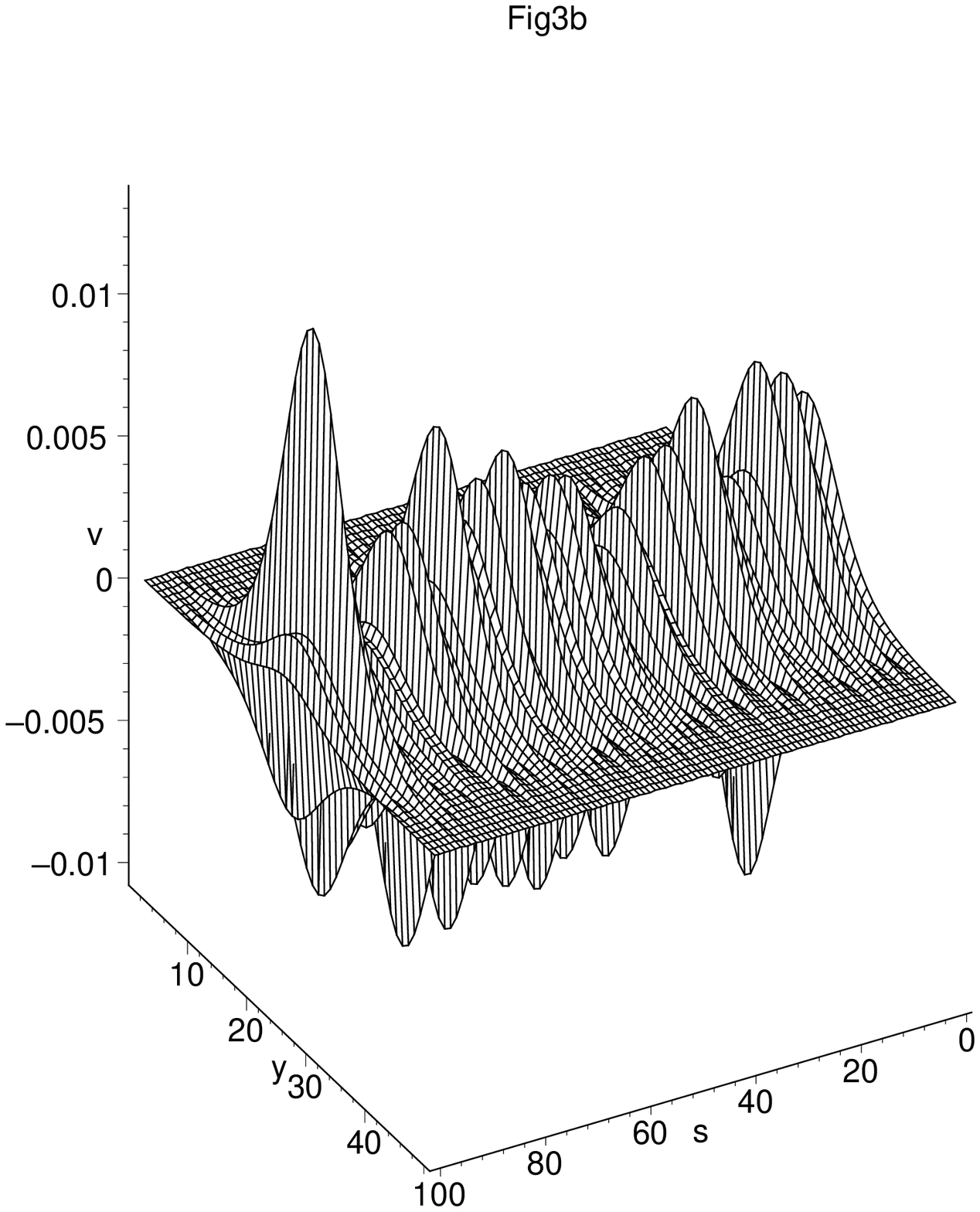}
\caption{\footnotesize (a) chaotic solution of the Lorenz system.
(b) The related chaotic line soliton solution of the 2DDLWE for
the field $v$.}
\end{figure}
Obviously, Fig. 3 shows us that when the parameters of (33) are
located at the chaotic regions, the corresponding solution becomes
a chaotic straight line soliton solution which is localized in $y$
direction and chaotic in $s$ direction.

\subsection{Space periodic and chaotic solutions}

From (28) we know that in some cases, the function $q$ may be an
arbitrary function of $y$, so we may also select it as a solutions
of the Lorenz system (33) with the replacement of the independent
variable $s\rightarrow y$. When the function $q=q(y)$ and
$w=w(s)=w(x+b(c+1)t)$ are all the solutions of the Lorenz system,
we can obtain many types of solutions which are periodic or
chaotic in both directions.

Fig.4 is a plot of a periodic solution of the 2DDLWE which has
periodic property in both directions while $q(y)$ and $w(s)$ are
chosen as both the solutions of the Lorenz system (33) with (45).
\input epsf
\begin{figure}
\epsfxsize=7cm\epsfysize=5cm \epsfbox{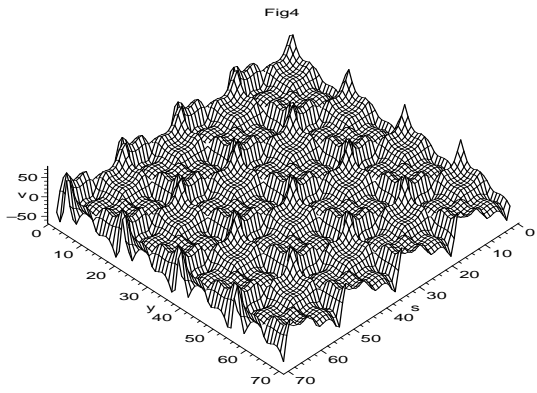}
{\footnotesize \caption{\footnotesize Periodic two solution of the
2DDLWE in both direction for the field $v$.}}
\end{figure}

Fig.5 is a plot of an exact solution of the 2DDLWE which is
periodic in $y$ direction and chaotic in $x$ direction. The
corresponding solutions for $q$ and $w$ are all determined by (33)
but with different parameters (45) and (47) respectively.
\input epsf
\begin{figure}
\epsfxsize=7cm\epsfysize=5cm \epsfbox{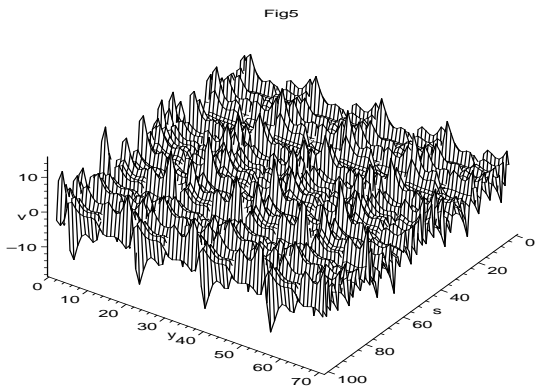}
\caption{\footnotesize Plot of the exact solution of the 2DDLWE
for the field which is periodic in $y$ direction and chaotic in
$x$ direction.}
\end{figure}

Fig.6 shows a chaotic solution of the 2DDLWE in both directions.
The related solutions for $q$ and $w$ are all determined by (33)
with the same parameters (47).
\input epsf
\begin{figure}
\epsfxsize=7cm\epsfysize=5cm \epsfbox{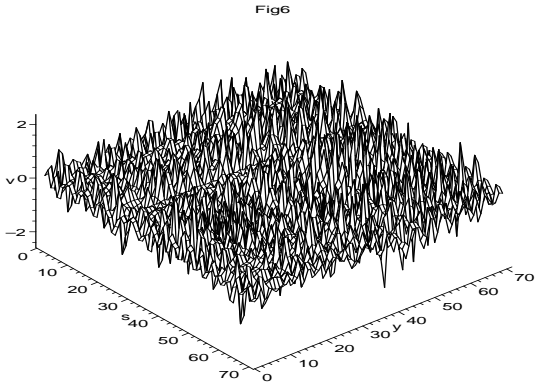}
\caption{\footnotesize A typical solution of the 2DDLWE which is
chaotic in both directions.}
\end{figure}
\section{Summary and discussions}

In summary, the CK's direct similarity reduction ansatz is
extended to a much more general form. Using the general reduction
ansatz, one may obtain various new lower dimensional reduction
equations including many turbulence and chaotic systems. Taking
the 2DDLWE as a concrete example, and a slightly special reduction
ansatz with three order derivatives of the reduction field, we
obtain a general three order quasi-linear equation, which includes
the KdV, MKdV, KdV-Burgers and the generalized Lorenz system as
special examples, as a special reduction of the 2DDLWE. The known
KS system and other types of higher order models may also be
obtained from the reductions of the 2DDLWE.

The reductions (30), (32) and (37) are known as some typical
turbulence and chaotic systems while the 2DDLWE is known as an IST
integrable model. The reason why some lower dimensional turbulence
and chaotic systems can be reduced from a higher dimensional
integrable (under some particular meanings) model is that for a
higher dimensional integrable model, some types of lower
dimensional arbitrary functions do enter into its general
solution.

Using the solutions of the lower dimensional models we may obtain
many kinds of new solutions for the 2DDLWE. Especially, using the
numerical solutions of the Lorenz systems, some types of periodic
line solitons, chaotic line solitons, periodic-periodic solution,
periodic-chaotic and chaotic-chaotic solutions of the 2DDLWE can
be obtained.

In Ref. $\cite{DS}$, using the variable separation approach, we
have also pointed out that the turbulence and chaotic systems can
be obtained from other ``integrable" models like the
Davey-Stewartson equations and the asymmetric
Nizhnik-Novikov-Veselov equation because of the entrance of the
lower dimensional arbitrary functions in the general solutions.
The more about the method and the effects of the reduced
turbulence system on the original model(s) are worthy of studying
further.

\vskip.2in The author is in debt to thanks the helpful discussions
with the professors Q. P. Liu, G-x Huang and C-p Sun. The work was
supported by the National Outstanding Youth Foundation of China,
the Research Fund for the Doctoral Program of Higher Education of
China and the Natural Science Foundation of Zhejiang Province of
China.

\end{document}